\documentstyle{aipproc}
\input epsf

\begin{document}
\title{Magnetohydrodynamic turbulence in warped accretion discs}
\author{Ulf Torkelsson$^*$, Gordon I. Ogilvie$^{\dagger}$,
Axel Brandenburg$^\ddagger$, James E. Pringle$^{\dagger}$, 
\AA ke Nordlund$^\|$ \and Robert F. Stein$^\P$}
\address{$^*$Chalmers University of Technology/G\"oteborg University,
Department of Astronomy and Astrophysics, S-412 96 Gothenburg, Sweden\\
$^{\dagger}$Institute of Astronomy, 
Madingley Road, Cambridge CB3 0HA, United Kingdom\\
$^\ddagger$Department of Mathematics, University of Newcastle upon Tyne, NE1 
7RU, United Kingdom \and Nordita, Blegdamsvej 17, DK-2100 Copenhagen \O, 
Denmark\\
$^\|$Theoretical Astrophysics Center, Juliane Maries Vej 30, DK-2100 
Copenhagen \O, Denmark \and Copenhagen University Observatory, 
Juliane Maries Vej 30, DK-2100 \\
$^\P$Department of Physics and Astronomy, Michigan State University, 
East Lansing, MI 48824, USA}
\maketitle
\begin{abstract}
Warped, precessing accretion discs appear in a range of astrophysical
systems, for instance the X-ray binary Her X-1 and in the active nucleus
of NGC4258.  In a warped accretion disc there are horizontal pressure
gradients that drive an epicyclic motion.  We have studied the interaction
of this epicyclic motion with the magnetohydrodynamic turbulence in
numerical simulations.  We find that the turbulent stress acting on the
epicyclic motion is comparable in size to the stress that drives the
accretion, however an important ingredient in the damping of the epicyclic
motion is its parametric decay into inertial waves.  
\end{abstract}

\section{Introduction}

Warped accretion discs have been a part of the astronomical vocabulary since
the discovery of the 35-day cycle of the X-ray binary Her X-1 
\cite{tananbaum,katz,roberts}.  Incidentally Her X-1 was the first 
occulting X-ray pulsar for which an optical counterpart was found as
discussed by Neta Bahcall at the 6th Texas Symposium in 1972
\cite{bahcall}.
Later on warped accretion discs have been
found in a multitude of systems.  In later years one of the most interesting
examples has been the maser source in the active galactic nucleus of
NGC 4258 \cite{miyoshi}.

While the warped accretion disc offered a simple interpretation of the
observations, it was not easy to describe it theoretically at the
hydrodynamic level.  The problems have been both to explain the
excitation mechanism of the warp and its coherence.  Pringle \cite{pringle}
showed that the radiation pressure from the central radiation source
may produce a warp in the outer disc, and Schandl \& Meyer \cite{SM}
described a similar mechanism in which the irradiation produces a wind,
which in its turn excites a warp.

A crucial condition for any of these mechanisms to work is that the
tendency of the disc to straighten itself must be sufficiently weak.
There are two different forces that strive to produce a flat disc.
Firstly there is the usual viscous stress due to the local turbulence,
which is also driving the accretion, but in general it is 
insignificant compared to the hydrodynamic stress due to the epicyclic
shear flow which is driven by the warp itself \cite{PP,PL}.
The amplitude of the epicyclic motion is inversely proportional to
the ordinary turbulent viscosity.  For that reason the hydrodynamic stress 
due to the epicyclic motion will also be inversely proportional to 
the turbulent viscosity, and the time scale for flattening the disc will
be anomalously short compared to the ordinary viscous time scale.

The warping instability therefore requires a mechanism that can damp
the epicyclic motion much more efficiently than it transports the 
angular momentum in the radial direction.  That would for instance be
the case if the turbulent viscosity was strongly anisotropic.  The
intention of this paper is to estimate how anisotropic the turbulent 
viscosity is and to check whether there are any other mechanisms that 
can limit the amplitude of the epicyclic motion.
We describe the numerical model that we use for these estimates in
Sect. 2.  There are then two ways in which we have studied the 
interaction between the turbulence and the epicyclic motion.  Firstly
we have studied the free decay of an epicyclic motion (Sect. 3) and 
secondly we have studied the motion that results from a radial forcing
(Sect. 4).  We discuss and summarize our results in Sect. 5.

\section{The mathematical model}

We solve the magnetohydrodynamical equations in a Keplerian shearing
box \cite{BNST,HGB,MS}.  Our units are chosen such that $H = GM = \mu_0 = 1$, 
where $G$ is
the gravitational constant, $M$ the mass of the central object, and
$H$ is the Gaussian scale height of the shearing box.  The density
distribution assuming isothermality
can then be written as $\rho = \rho_0 e^{-z^2/H^2}$,
where we put $\rho_0 = 1$.
The physical size of the shearing box is $L_x : L_y : L_z =
1 : 2\pi : 4$, and the box is positioned such that $x$ and $z$ vary between
$\pm\frac{1}{2} L_x$, and $\pm\frac{1}{2} L_z$, respectively, while 
$y$ goes from 0 to $L_y$, and the distance of the origin to the central
object, $R_0 = 10$.
This gives the orbital period $T_0 = 199$, and the mean internal energy
$e_0 = 7.4\,10^{-4}$.  To stop the box from heating up we add a cooling
function
\begin{equation}
  Q = - \sigma_{\rm cool}\left(e-e_0\right),
\end{equation}
where $\sigma_{\rm cool}$ is the cooling rate, which typically
corresponds to a time scale of 1.5 orbital periods.
Our boundary conditions are (sliding) periodic in the ($x$-) $y$-direction.
In the $z$-direction they are impenetrable and stress-free for the
velocity, and acts as a perfect conductor with respect to the magnetic
field.

We start the simulations from a snapshot from a previous simulation
in which the magnetohydrodynamic turbulence is already fully developed.
In the first set of simulations we then add a radial velocity of the form
$u_x = u_0\sin(\pi z/L_z)$, which will have the time evolution of an
epicyclic motion.  In the second set of simulations, we do not modify
the velocity field of the initial snapshot, but rather add a radial
forcing term to the equation of motion. 
This forcing gives an acceleration with a harmonic time-dependence on the
orbital time scale and the same $z$-dependence as the velocity above.

\begin{figure}
\epsfxsize=14cm
\epsfbox{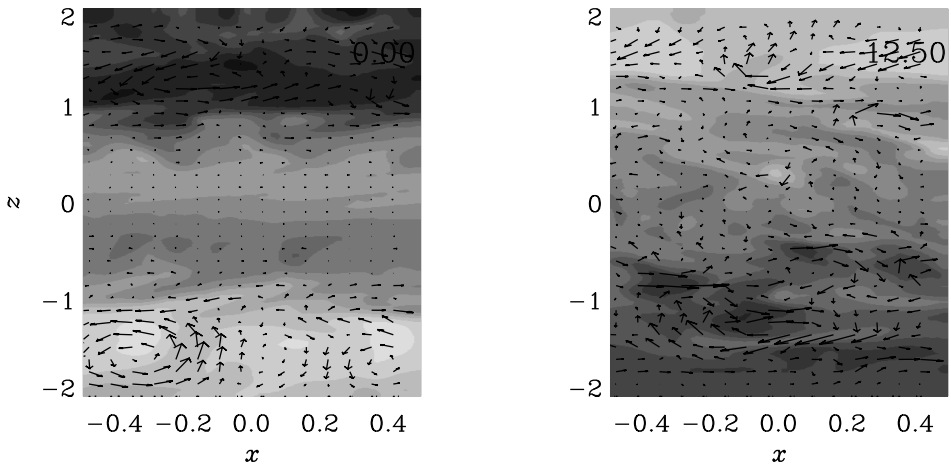}
\caption{The magnetic field in a meridional cut of the simulation at the
beginning of the simulation (left) and after 12.5 orbital periods (right).
The toroidal field is plotted using a grey scale and the poloidal field
as vectors.  Note that the toroidal field has been reversed between 
the two images.}
\label{snaps}
\end{figure}

\section{The free decay of an epicyclic motion}

In the first set of simulations we add an epicyclic motion to the
initial state of the simulation, and then follow the decay of the 
epicyclic motion.  These simulations have previously been described
in \cite{torkel}.  With a weak epicyclic motion, its maximum Mach number
is initially 0.38, it is difficult to follow the evolution of the
epicyclic motion as it is comparable in size to the turbulent
velocities.  When the amplitude is increased to a Mach number of 3.3,
we can distinguish two stages in the damping.  After a brief period of
essentially no damping the epicyclic velocity quickly drops by a factor
of 2 to 3.  This damping is followed by an extended phase of exponential
decay with a time scale of 25 orbital periods.  The time scale of
the exponential decay can be translated to a Shakura-Sunyaev \cite{SS}
$\alpha$-parameter of 0.006, which is within a factor of two of the
values usually derived from turbulence simulations, e.g. \cite{BNST,stone}.
The preceding rapid damping is a new phenomenon though, which we
interpret in terms of that the epicyclic motion is decaying to 
inertial waves via a parametric instability \cite{GGO}.
We also note that the toroidal magnetic field in the shearing box
reverses its direction during the damping of the strong epicyclic
motion (Fig. \ref{snaps}).

\begin{figure}
\epsfxsize=14cm
\epsfbox{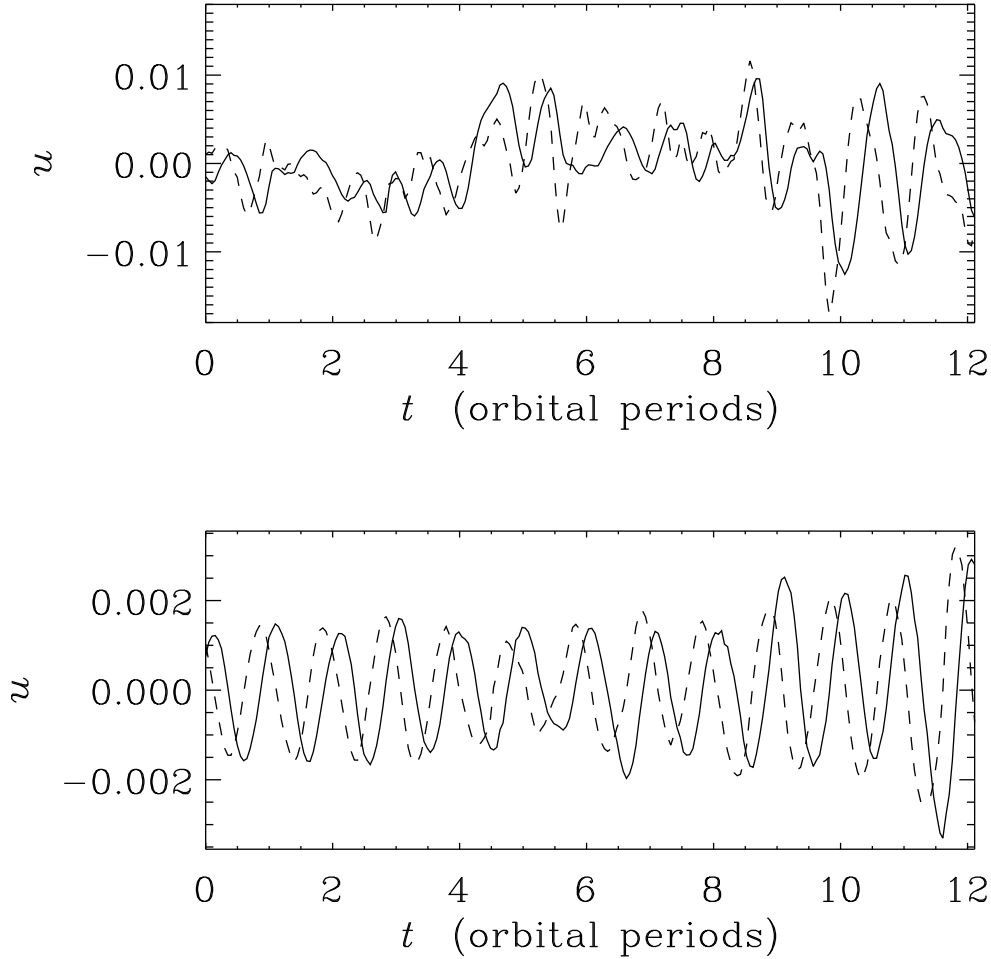}
\caption{$\langle u_x\rangle$ (solid line) and $2 \langle u_y\rangle$ (dashed
line) as functions of time at $z = 1.17$ (top) and $z = -0.25$ (bottom).}
\label{epic}
\end{figure}

\section{Driven epicyclic motion}

The dynamics in a real accretion disc is significantly different from
the case we have studied above.  In reality the epicyclic motion will 
be driven by the radial pressure gradient that is set up by the warp.
To mimic this 
we have carried out a new set of simulations in which we drive the
epicyclic motion by adding a time-periodic radial force to the equation of
motion. 

We plot the horizontally averaged velocities $\langle u_x\rangle$
and $\langle u_y\rangle$ in Fig. \ref{epic}.  The epicyclic motion is
in this case comparable to that of Run 1b in \cite{torkel}.  The
results are similar in the two cases, and the epicyclic motion is
difficult to distinguish at $z = 1.17$ due to the effect of the
turbulent stresses, while it is easily distinguishable at $z = -0.29$.

\section{Conclusions}

In this paper we have studied the dynamics of an epicyclic flow in a 
Keplerian shear flow.  In our simulations we find two damping mechanisms
for the epicyclic motion.
The turbulent stresses can damp the motion in a way which can be
described in terms of a turbulent viscosity comparable in strength to
that driving the radial angular momentum transport, but a 
sufficiently fast epicyclic motion can lose significant amounts of
energy by exciting inertial waves through a parametric instability.

\section*{Acknowledgments}

Computer resources from the National Supercomputer Centre at Link\"oping
University are gratefully acknowledged.  UT is supported by the Swedish
Research Council (formerly the Natural Sciences Research Council, NFR), and
RFS is supported by NASA grant NAG5-4031.
This work was supported in part by the 
Danish National Research Foundation through its establishment of the
Theoretical Astrophysics Center (\AA N).

\end{document}